\documentclass[aps,prl,amsmath,amssymb,twocolumn, reprint]{revtex4-1}

\usepackage{graphicx}
\usepackage{dcolumn}
\usepackage{footmisc}

\usepackage{bm}
\usepackage{hyperref}
\usepackage{xcolor}
\usepackage{float}
\begin{document}
\setlength{\abovedisplayskip}{5pt}
\setlength{\belowdisplayskip}{5pt}
\setlength{\abovedisplayshortskip}{5pt}
\setlength{\belowdisplayshortskip}{5pt}

\preprint{}

\title{Probing Light Dark Matter through Cosmic-Ray Cooling in Active Galactic Nuclei}

\author{Gonzalo Herrera\textsuperscript{1,2}, Kohta Murase\textsuperscript{3,4,5}}
\affiliation{
\textsuperscript{1}Physik-Department, Technische Universit\"at M\"unchen, James-Franck-Stra\ss{}e, 85748 Garching, Germany,\\
\textsuperscript{2}Max-Planck-Institut f\"ur Physik (Werner-Heisenberg-Institut), F\"ohringer Ring 6,80805 M\"unchen, Germany,\\
\textsuperscript{3}Department of Physics; Department of Astronomy \& Astrophysics; Center for Multimessenger Astrophysics, Institute for Gravitation and the Cosmos, The Pennsylvania State University, University Park, PA 16802, USA,\\ 
\textsuperscript{4}School of Natural Sciences, Institute for Advanced Study, Princeton, NJ 08540, USA,\\
\textsuperscript{5}Center for Gravitational Physics and Quantum Information, Yukawa Institute for Theoretical Physics, Kyoto University, Kyoto, Kyoto 606-8502, Japan
}

\begin{abstract}
Recent observations of high-energy neutrinos from active galactic nuclei (AGN), NGC 1068 and TXS 0506+056, suggest that cosmic rays (CRs) are accelerated in the vicinity of the central supermassive black hole and high-energy protons and electrons can cool efficiently via interactions with ambient photons and gas. The dark matter density may be significantly enhanced near the black hole, and CRs could lose energies predominantly due to scatterings with the ambient dark matter particles. We propose CR cooling in AGN as a new probe of dark matter-proton and dark matter-electron scatterings. Under plausible astrophysical assumptions, our constraints on sub-GeV dark matter can be the strongest derived to date. Some of the parameter space favored by thermal light dark matter models might already be probed with current multimessenger observations of AGN.
\end{abstract}

\maketitle


\begin{figure}
    \centering
    \includegraphics[width=0.5\textwidth]{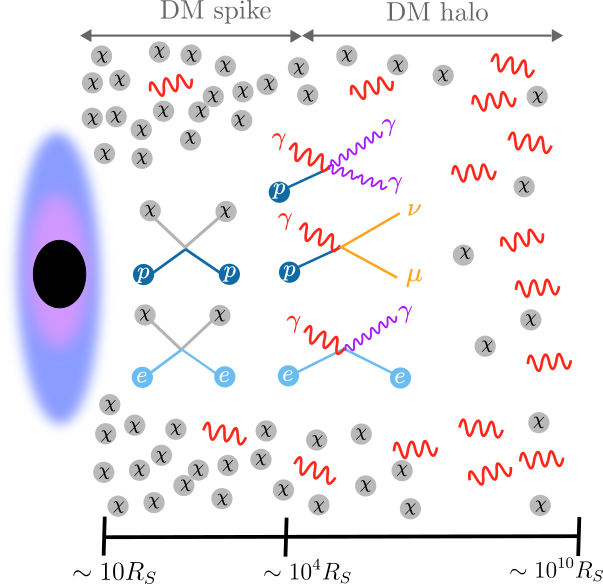}
    \caption{Schematic picture of dark cooling of CRs due to elastic scatterings with DM particles in AGN. High-energy protons and electrons may traverse a high density of DM particles more efficiently than standard cooling processes involving neutrino and photon emission. 
    }
    \label{fig:diagram}
\end{figure}
	
\begin{table*}[t!]
		\begin{center}
			\begin{tabular}{l|ccccccc}
				\hline
				& $R_{\rm em}$ &  $M_{\rm BH}$ & $t_{\rm BH}$ & $r_0$ & $\langle\sigma v \rangle$/$m_{\rm DM}$ & $\langle \rho_{\rm DM} \rangle$  \\
				\hline
				\toprule
				 NGC 1068 (I) & 30 $R_S$  & (1-2) $\times 10^{7}M_{\odot}$ &  $10^{10}$ yr & 10 kpc & 0 & $5\times10^{18}$ GeV/cm$^{3}$   \\
				 NGC 1068 (II) & 30 $R_S$ & (1-2) $\times 10^{7}M_{\odot}$ &  $10^{10}$ yr & 10 kpc & $10^{-31}$cm$^{3}$s$^{-1}$/GeV & $4\times10^{13}$ GeV/cm$^{3}$      \\
				 TXS 0506+056 (I) & $10^4 R_S$ &  (3-10) $\times 10^{8}M_{\odot}$ &  $10^9$ yr & 10 kpc & 0 & $8\times10^{12}$ GeV/cm$^{3}$  \\
				 TXS 0506+056 (II) & $10^4 R_S$ & (3-10) $\times 10^{8}M_{\odot}$ &  $10^9$ yr & 10 kpc &$10^{-28}$cm$^{3}$s$^{-1}$/GeV & $4\times10^{11}$ GeV/cm$^{3}$   \\
                \hline
			\end{tabular}
		\end{center}
		\caption{Relevant parameters considered in this work for NGC 1068 and TXS 0506+056, for two different sets of assumptions dubbed (I) and (II). Here $R_{\rm em}$ represents the distance of the emission region from the central SMBH in NGC 1068 (TXS 0506+056), $M_{\rm BH}$ shows the SMBH mass and its uncertainty, $t_{\rm BH}$ is the black hole age, $r_0$ is the scale radius of the galaxy, $\langle\sigma v \rangle$/$m_{\rm DM}$ denotes the assumed values of the effective DM self-annihilation cross section, and $\langle \rho_{\rm DM} \rangle$ is the average density of DM particles within $R_{\rm em}$.}
		\label{tab:AGN}
	\end{table*}

The presence of dark matter (DM) in galaxies and clusters of galaxies is well established by astrophysical and cosmological observations \cite{Bertone:2016nfn}. However, its particle nature remains unknown \cite{Bertone:2004pz}. A variety of experiments have aimed to detect DM particles via their scatterings off nuclei and/or electrons at Earth-based detectors, setting strong upper limits on the interaction strength of DM particles with masses in the GeV scale, but leaving the sub-GeV region of the parameter space yet largely unconstrained \cite{Goodman:1984dc,Bernabei:2007gr,MarrodanUndagoitia:2015veg,Essig_2012,Essig:2022dfa}. 
Historically, DM fermions with sub-GeV masses were disfavored by the cosmological bound 
\cite{HUT197785,Lee:1977ua}. 
However, large parameter space still remains unexplored in more complicated but well-motivated scenarios \cite{Boehm:2003hm,Hall:2009bx, Hochberg:2014dra}.

Different approaches have been proposed to extend the sensitivity reach of direct detection experiments for sub-GeV DM. Some of these consider a boosted component of DM particles reaching the Earth, via gravitational effects, or via scatterings with protons, electrons or neutrinos in different astrophysical environments (\textit{e.g.,} Refs.~\cite{Baushev:2012dm, Behroozi_2013, Besla:2019xbx, Herrera:2021puj, Herrera:2023fpq, Smith-Orlik:2023kyl, Bringmann:2018cvk, Ema:2018bih, Alvey:2019zaa, Wang:2021jic, Agashe:2014yua, Kim:2016zjx, Das:2021lcr, Cappiello:2022exa, Arguelles:2022fqq,Lin:2022dbl, Maity:2022exk}).

Active galactic nuclei (AGN) are promising sources of high-energy protons and electrons. While the dominant acceleration mechanism of these cosmic rays (CRs) is still under debate, modeling of multimessenger data have placed important constraints on not only energetics of CR production but also the emission region of the observed neutrinos that can be produced either via inelastic $pp$ collisions or $p\gamma$ interactions \cite{Murase:2013rfa,Winter:2013cla,Murase_2016}. 
For example, observations of high-energy neutrinos and gamma rays from NGC 1068 \cite{IceCube:2022der,Ajello:2023hkh} suggest that the neutrino production occurs in the vicinity of the supermassive black hole (SMBH) at $R_{\rm em }\lesssim 100 R_S$ (where $R_S$ is the Schwarzschild radius), which is consistent with theoretical models \cite{Murase:2019vdl,Kheirandish:2021wkm,Eichmann:2022lxh,Inoue:2019yfs,Inoue:2022yak}, and the required proton luminosity is at least $\sim10\%$ of the intrinsic X-ray luminosity \cite{Murase:2022dog,Blanco:2023dfp}. 
Another neutrino source candidate, TXS 0506+056 \cite{IceCube:2018dnn,IceCube:2018cha,MAGIC:2018sak,Keivani:2018rnh,Padovani_2019}, is known to be a jet-loud AGN, and the observed spectral energy distribution in photons is mostly explained by synchrotron and inverse-Compton emission from CR electrons 
\cite{Keivani:2018rnh,Murase:2018iyl,Gao:2018mnu,Cerruti_2018,Zhang:2019htg, Petropoulou_2020}, and the proton luminosity required by IceCube observations may even exceed the Eddington luminosity \cite{Murase:2018iyl,Gao:2018mnu}. 

In this work, we propose CRs produced in AGN as a new, unique probe of DM-proton and DM-electron scatterings through their multimessenger observations. Given that emission regions of neutrinos and gamma rays are constrained to be near the SMBHs, CRs also need to traverse the DM spike around the central SMBH. If such additional cooling beyond the SM (BSM) was too strong, CR energy losses are modified so that the required CR luminosity would be larger, and the neutrino and photon spectra could even be incompatible with the observations. 
Our work is different from previous studies on AGN probes of the DM scatterings with protons and electrons, which focused either on the boosted flux of DM particles from the source at Earth \cite{Wang:2021jic,Granelli_2022,Bhowmick:2022zkj}, or on the spectral distortions in the gamma-ray flux induced by CR scatterings off DM particles \cite{Bloom_1998, Gorchtein_2010, Hooper:2018bfw, Cappiello:2018hsu, Cermeno:2022rni, Ambrosone:2022mvk}. Instead we focus on the impact of the DM-proton and DM-electron scatterings on the neutrino and photon fluxes or the CR power, considering for the first time the cooling of protons and electrons in the inner regions of the AGN, where a DM spike is likely to be formed.

{\it DM distribution in AGN---.}The adiabatic growth of black holes may form a spike of DM particles in their vicinity \cite{1972GReGr...3...63P,Quinlan:1994ed,Gondolo_1999, Ullio:2001fb}. An initial DM profile of the form $\rho (r) = \rho_0 (r/r_0)^{-\gamma}$ evolves into:
\begin{align}
	\rho_{\rm sp}(r) = \rho_{R} \, g_{\gamma}(r)\, \bigg(\frac{R_{\rm sp}}{r}\bigg)^{\gamma_{\rm sp}}\;,
\end{align}
where $R_{\rm sp}=\alpha_{\gamma}r_0(M_{\rm BH}/(\rho_{0}r_{0}^{3}))^{\frac{1}{3-\gamma}}$ is the size of the spike, with numerical values $\alpha_\gamma$ provided in Ref.~\cite{Gondolo_1999}. The spike slope is by $\gamma_{\rm sp}=\frac{9-2\gamma}{4-\gamma}$. Further, $g_{\gamma}(r) \approx (1-\frac{4R_{\rm S}}{r})$, while $\rho_{\rm R}$ is a normalization factor, chosen to match the density profile outside of the spike, $\rho_R=\rho_{0}\, (R_{\rm 
sp}/r_0)^{-\gamma}$. This density profile vanishes at $4 R_{\rm S}$, which is a conservative approximation \cite{Ferrer:2017xwm, Ferrer:2022kei}.

\begin{figure*}[t!]
	\centering
	\includegraphics[width=0.49\textwidth]{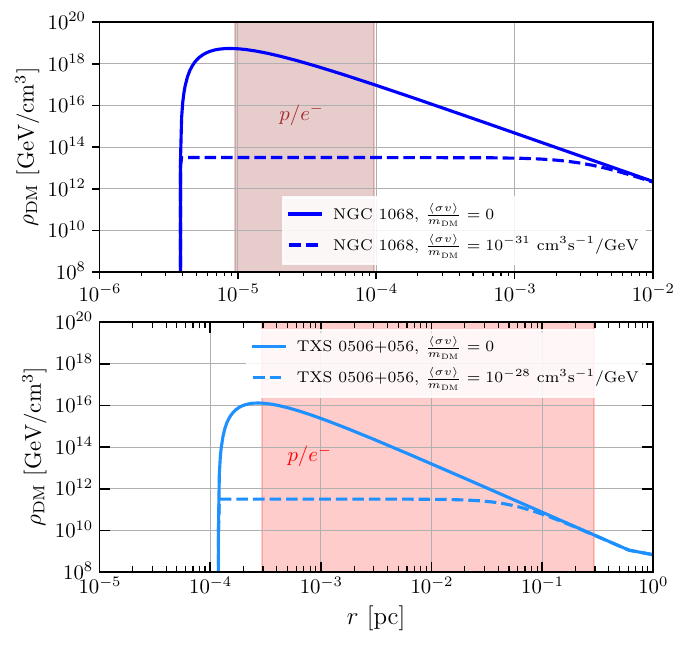}
	\includegraphics[width=0.49\textwidth]{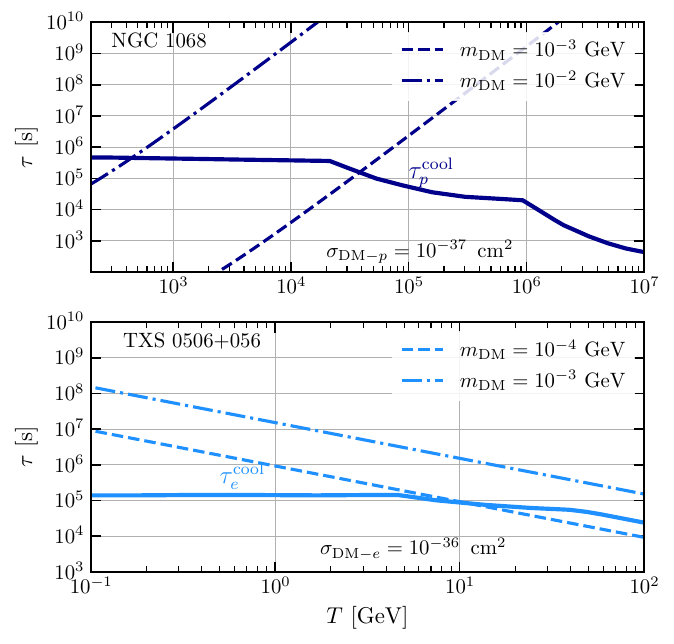}
	\centering
	\caption{{\it Left panel}: DM distribution around the SMBHs of TXS 0506+056 and NGC 1068, for different values of the DM self-annihilation cross section over its mass. The red (brown) shaded region indicates the region where the production of high-energy particles is expected to take place in TXS 0506+056 (NGC 1068) 
    {\it Right panel}: In solid lines, we show the cooling timescales of protons (electrons) in NGC 1068 (TXS 0506+056). In dashed and dot-dashed lines we show the timescales due to elastic DM proton and DM-electron scatterings \cite{Murase:2022dog, Keivani:2018rnh}.
    }
	\label{fig:DMspike}
\end{figure*}

We consider that the initial DM distribution follows an NFW profile \cite{Navarro:1996gj,Navarro:1995iw}, with $\gamma=1$, resulting in a spike with $\gamma_{\rm sp}=7/3$ and $\alpha_{\gamma}= 0.122$. The masses of the central SMBHs of the two AGN considered in this work are given in Table \ref{tab:AGN}. For the scale radius of both galaxies, we take $r_0=10$ kpc. Finally, the normalization $\rho_0$ is determined by the uncertainty on the SMBH mass \cite{Gorchtein_2010,Lacroix_2017}, also provided in Table \ref{tab:AGN}. We have checked that this criteria yields masses of the DM halo compatible with those expected from universal relations between SMBH and galactic bulge masses \cite{DiMatteo:2003zx,Ferrarese:2002ct}. 
We use Eq.~(8) of Ref.~\cite{DiMatteo:2003zx}. 

If the DM particles self-annihilate, the maximal DM density in the inner regions of the spike is saturated to $\rho_{\text {sat}}= m_{\rm DM} /(\langle\sigma v \rangle t_{\mathrm{BH}})$, where $\langle \sigma v \rangle$ is the velocity averaged DM self-annihilation cross section, and $t_{\rm BH}$ is the SMBH age. For TXS 0506+056 (NGC 1068), we take the value $t_{\rm BH}=10^9$ ($10^{10}$) yr \cite{Granelli_2022,Piana_2020}. 
Furthermore, the DM spike extends to a maximal radius $R_{\rm sp}$, beyond which the DM distribution follows the initial NFW profile. The DM density profile therefore reads \cite{Gondolo_1999, Lacroix_2015, Lacroix_2017})
\begin{align}\rho(r)= \begin{cases} 
		0 & r\leq 4R_{\rm S}, \\[5pt]
		\frac{\rho_{\rm sp}(r)\rho_{\rm sat}}{\rho_{\rm sp}(r)+\rho_{\rm sat}} & 4R_{\rm S}\leq r\leq R_{\rm sp}, \\[3pt]
		\rho_{0}\left(\frac{r}{r_0}\right)^{-\gamma} \left(1+\frac{r}{r_0}\right)^{-(3-\gamma)} & r\geq R_{\rm sp} .
	\end{cases}
	\label{eq:spike_profile}
\end{align}
The DM profiles in TXS 0506+056 and NGC 1068 are shown in the left panel of Fig.~\ref{fig:DMspike} for two values of $\langle \sigma v\rangle/m_{\rm DM}$ allowed for sub-GeV DM \cite{Lin:2011gj, Essig:2013goa, Cirelli:2020bpc}. We find that the DM density is extremely high in the region where high-energy particles are produced.

{\it BSM Cooling of CRs in AGN---.}Neutrinos and photons from AGN can be explained by emission from high-energy protons and electrons through purely SM mechanisms. Energy-loss mechanisms include scatterings with other SM particles in the plasma or synchroton radiation as well as adiabatic losses. In addition, there are escape losses due to advection or diffusion via magnetic fields. 
The presence of DM coupling to protons and electrons in the vicinity of SMBHs would introduce additional scattering timescales, leading to the suppression of the observed neutrino and gamma-ray fluxes in certain energy ranges, if the BSM cooling timescales of CRs were shorter than the standard cooling timescales. 
For example, at $m_{\rm DM} \sim 10^{-3}~$GeV, the currently-allowed maximum DM-proton cross section stems from CR boosted DM at the Super-Kamiokande experiment \cite{Super-Kamiokande:2022ncz}, with a value of $\sigma_{{\rm DM}-p} \sim 10^{-35}$~cm$^2$. As discussed previously, the average density of asymmetric DM particles in the corona of NGC 1068 is $\langle \rho_{\rm DM} \rangle \sim 5 \times 10^{18}$~GeV/cm$^3$. Thus, if the corresponding cross section for CR protons is comparable to $\sigma_{{\rm DM}-p}$ (although this is not the case in general), the BSM cooling timescale for the currently allowed values in the literature is $\tau_{{\rm DM}-p} \sim 1/(\langle n_{\rm DM} \rangle \sigma_{{\rm DM}-p} c) \sim 7 \times 10^{3}~{\rm s}$, which is well below the proton cooling time inferred by observations of NGC 1068. This simple estimate clearly suggests that CRs in AGN can provide a powerful probe of these interactions.

More quantitatively, the BSM cooling timescale due to elastic DM scattering off CRs is given by \cite{Ambrosone:2022mvk}
\begin{equation}\label{eq:tau_el}
    \tau^{\rm el}_{\text{DM}-i}=\Bigg[-\frac{1}{E}\bigg(\frac{dE}{dt}\bigg)_{\text{DM}-i}\Bigg]^{-1},
\end{equation}
with
\begin{equation}\label{eq:dmenergyloss}
    \left(\frac{dE}{d t}\right)_{\text{DM}-i} = -\frac{\langle \rho_{\rm DM}\rangle}{{m_{\rm DM}}}\,\int_{0}^{T^{\rm max}_{\rm DM}} dT_{\rm DM}\, T_{\rm DM} \frac{d\sigma_{\text{DM}-\text{CR}i}}{dT_{\rm DM}} \,,
\end{equation}
where $\langle \rho_{\rm DM} \rangle$ is the average density of DM particles in the region of CR production. See Table~\ref{tab:AGN} for the specific values that we use for NGC 1068 and TXS 0506+056. 
Also, $d\sigma_{\text{DM}-\text{CR}i}/dT_{\rm DM}$ is the differential elastic DM-proton or DM-electron cross section, $T^{\rm max}_{\rm DM}$ is the maximal allowed value for $T_{\rm DM}$ in a collision with a particle $i$ with kinetic energy $T = E-m_i$, which is
\begin{equation}\label{eq:energygain}
    T^{\rm max}_{\rm DM} = \frac{2T^2 + 4m_i T}{m_{\rm DM}}\left[\left(1+\frac{m_i}{m_{\rm DM}}\right)^2 + \frac{2 T}{m_{\rm DM}}\right]^{-1} \,.
\end{equation}
We consider fermionic DM which elastically interacts with protons and electrons via a heavy scalar mediator. The differential cross section reads \cite{Bondarenko:2019vrb}
\begin{equation}\label{eq:differentialcrosssection}
    \frac{d\sigma_{\text{DM}-\text{CR}i}}{dT_{\rm DM}} = \frac{\sigma_{{\rm DM}-i}}{ T^{\rm max}_{\rm DM}} \frac{F_{i}^{2} (q^2)}{16 \,\mu_{{\rm DM} -i}^2\,s} \,  (q^2 + 4m^2_i)(q^2 + 4m^2_{\rm DM})\,,
\end{equation}
where $\sigma_{{\rm DM}-i}$ is the DM-proton or DM-electron cross section at the zero center-of-mass momentum, $\mu_{{\rm DM}-i}$ is the reduced mass, $s$ is the square of center-of-mass energy, and $q^2=2m_{\rm DM} T_{\rm DM}$ is the momentum transfer of the process. The quantity $F_i$ is either the proton form factor \cite{Angeli:2004kvy}, or equal to 1 for electrons.

In the right panel of Fig.~\ref{fig:DMspike}, the solid lines represent the total standard energy-loss timescales as a function of energy for protons in NGC 1068 \cite{Murase:2022dog} and for electrons in TXS 0506+056 \cite{Keivani_2018}. 
CR protons in NGC 1068 are almost depleted, and the dominant cooling mechanisms at increasing energies are inelastic $pp$ interactions, Bethe-Heitler pair production, and $p\gamma$ interactions \cite{Murase:2019vdl}. 
CR protons do not cool efficiently in TXS 0506+056, and the fate is governed by a dynamical timescale of $\sim 10^5$~s in the SMBH frame \cite{Keivani_2018}. 
For electrons in TXS 0506+056, the dominant cooling mechanisms are inverse Compton scattering and synchroton radiation. The breaks in the solid lines of the plots reflect the energies at which the transition of dominant processes occurs. 

For comparison, we also show BSM cooling timescales due to elastic DM scatterings with protons and electrons. The dotted-dashed line corresponds to values of the DM mass and cross section that would induce a contribution smaller than the proton and electron energy losses due to the SM processes. 
On the other hand, the dashed line shows values of the parameters that would induce larger energy losses than in the SM at relevant energies. It is important to point out that inelastic DM-proton scatterings are expected to dominate over the elastic channel at energies $E\gtrsim m_{p}^{2}/2m_{\rm DM}$. For simplicity, we restrict our analysis to the elastic channel.

\begin{figure*}[t!]
    \centering
	\includegraphics[width=0.475\textwidth]{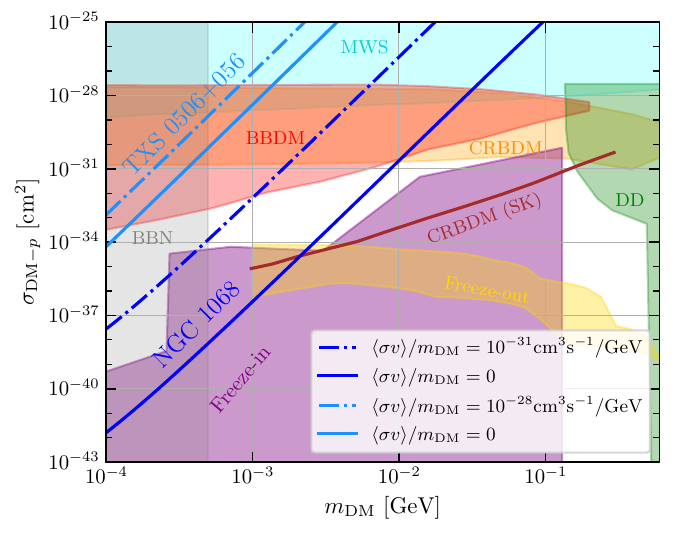}
	\includegraphics[width=0.475\textwidth]{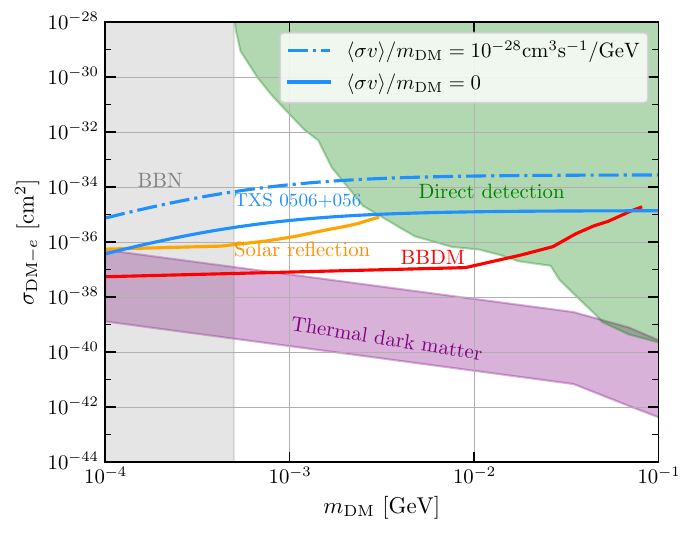}	
	\centering
	\caption{{\it Left panel:} Upper limits on the DM-proton cross-section vs DM mass, derived from the requirement that the required proton luminosity is substantially larger due to scatterings off DM particles in NGC 1068 and TXS 0506+056. Complementary constraints from different searches are shown for comparison. We also show values of the DM-proton scattering cross section that can produce the observed DM relic abundance via freeze-in (purple) \cite{Elor:2021swj,Bhattiprolu:2022sdd} and via freeze-out (gold) \cite{Smirnov:2020zwf,Parikh:2023qtk}.
    {\it Right panel:} Analogous upper limits on the DM-electron cross section, for TXS 0506+056. We also display values of the DM-electron cross section vs DM mass compatible with thermal production of light DM \cite{essig2023snowmass2021}. This band shows the range of values compatible with thermal production in different models (scalar dark matter \cite{Essig:2015cda}, asymmetric dark matter \cite{Lin:2011gj}, SIMPs \cite{Hochberg:2014dra, Hochberg:2014kqa} and ELDERs \cite{Kuflik:2015isi}). Note that we do not have constraints from NGC 1068 because the observed gamma rays may be purely hadronic \cite{Ajello:2023hkh}.
    }
	\label{fig:UpperLimitsMass}
\end{figure*}

For the purpose of constraining the interaction strength, we find for each $m_{\rm DM}$ the largest DM-proton (electron) cross section yielding a timescale equal or larger to the cooling timescales determined with models at relevant energies. In particular, we use
\begin{equation}\label{eq:criteria}
\tau^{\rm el}_{\text{DM}-i}\geq C \, \tau^{\rm cool}_{i}.
\end{equation}
The coefficient $C$ is a model dependent factor, and we use $C=0.1$ in this work. In other words, we find the maximum DM-proton (electron) cross section that would have an $\mathcal{O}(10)$ or less impact on the proton (electron) cooling timescale. This is reasonable and may be even conservative from the energetics requirement of neutrino-emitting AGN. For NGC 1068, the proton luminosity would be $10^{43}~{\rm erg}~{\rm s}^{-1}\lesssim L_p\lesssim L_X\lesssim10^{44}~{\rm erg}~{\rm s}^{-1}$~\cite{Murase:2019vdl,Murase:2022dog}, justifying $C\sim0.1-1$. For TXS 0506+056, the proton luminosity in the single-zone model already violates the Eddington luminosity $L_{\rm Edd}$ \cite{Murase:2018iyl}, so our choice is conservative. This is also reasonable for electrons because of $L_{e}\sim 8\times{10}^{47}~{\rm erg}~{\rm s}^{-1}\sim20 L_{\rm Edd}$ \cite{Keivani_2018}.    
In principle, if the CR acceleration mechanism is understood, spectral modification due to BSM cooling may allow us to improve constraints and $C\sim1$ is possible. For proton energies of interest, we use $10-300$~TeV for NGC 1068 \cite{IceCube:2022der}, which is required to match the IceCube data \cite{Murase:2022dog}. For protons in TXS 0506+056, we use $0.1-20$~PeV \cite{IceCube:2018cha}, and for electrons we use $50~{\rm GeV}-2~{\rm TeV}$, following Ref.~\cite{Keivani_2018}.  

Applying the condition of Eq.~(\ref{eq:criteria}) for NGC 1068 and TXS 0506+056, we set constraints on the DM-proton and DM-electron cross sections via a heavy mediator (see Fig.~\ref{fig:UpperLimitsMass}). 
The solid lines correspond to scenario (I), and the dashed lines correspond to scenario (II) (see Table~\ref{tab:AGN}).  We find that our constraints become stronger at lower DM masses, due to the fact that the number density of DM particles increases, and the cross section needed to induce energy losses becomes smaller. However, for protons the dependence of the constraint on the DM mass is more pronounced than for electrons, since the elastic DM-proton cross section decreases with reference to its maximum value for $E \gtrsim m_p^2/2 m_{\rm DM}$.

For comparison, we show complementary constraints from other methods. The green region is excluded by DM direct detection experiments \cite{CRESST:2019jnq, Amole_2016, Barak_2020, Aprile_2019, Emken:2017erx,Emken:2019tni, Mahdawi_2018}. The cyan region is constrained by Milky Way satellite galaxy counts \cite{Buen-Abad:2021mvc}, and the grey region is constrained by Big Bang nucleosynthesis \cite{Depta_2019, Giovanetti:2021izc}. The orange region is constrained by CR boosted DM at XENON1T \cite{Bringmann:2018cvk}, and the red regions are excluded when considering the blazar-boosted DM flux from TXS 0506+056 \cite{Wang:2021jic, Granelli_2022}. Finally, values above the brown line are constrained by CR boosted DM at the Super-Kamiokande experiment \cite{Super-Kamiokande:2022ncz}.
Further, for DM-electron scatterings, we include constraints from the solar reflection \cite{An_2018}, and the region of values where light thermal DM acquires its relic abundance via various mechanisms \cite{Hochberg:2014dra, Kuflik:2015isi,Essig_2017}.
From Fig.~\ref{fig:UpperLimitsMass}, one sees that our constraints for light DM coupling to protons are stronger than complementary bounds for $m_{\rm DM}\lesssim 10^{-3}-10^{-2}$~GeV. Additional constraints from colliders may also apply, see \textit{E.g} \cite{Daci:2015hca} for constraints on dark matter with masses above $m_{\rm DM} \gtrsim$ 0.1 GeV, assuming a mediator of mass $m_{\phi} \sim 1$ GeV and a direct coupling to quarks. If the mediator were more massive, or if it couples to gluons, those constraints could be weaker.

For DM-electron scatterings, our constraints are stronger than direct detection bounds at masses below $m_{\rm DM} \lesssim 5 \times 10^{-3}$~GeV. In addition, for DM-electron interactions, AGN data allows to probe the parameter space favored for DM models with $m_{\rm DM} \lesssim 10^{-4}$~GeV. Effects of different assumptions on the DM distribution and constraints in a concrete model of DM-proton interactions are discussed in the Supplementary Material.\\

{\it Summary and Discussion---.}
Recent multimessenger measurements of AGN have indicated that high-energy particles, in particular CR protons and secondary neutrinos, are produced in the vicinity of SMBHs. CR cooling could be significantly enhanced by BSM interactions with DM, thanks to a large DM density around the central SMBH. We demonstrated that neutrino-emitting AGN, NGC 1068 and TXS 0506+056, allow us to set strong constraints on sub-GeV DM coupled to protons and/or electrons. The new constraints on light DM coupling to protons are stronger than other complementary bounds for $m_{\rm DM}\lesssim 10^{-3}-10^{-2}$~GeV. For DM-electron scatterings, our constraints are stronger than direct detection bounds at masses below $m_{\rm DM} \lesssim 5 \times 10^{-3}$~GeV, which potentially allows us to probe the parameter space favored for thermal DM models with $m_{\rm DM} \lesssim 10^{-4}$~GeV.

Remarkably, our method based on CR cooling is unique and different from previous AGN constraints from boosted DM~\cite{Wang:2021jic} and those on neutrino-DM interactions~\cite{Kelly:2018tyg,Choi:2019ixb,Murase_2019,Alvey_2019,Cline:2022qld,Ferrer:2022kei,Cline:2023tkp}, in that the results are largely insensitive to CR and neutrino spectra. Regarding uncertainties in CR cooling timescales, we stress that our constraints are robust and conservative for NGC 1068. This is because the neutrino production efficiency of CR protons has to be nearly maximal to explain the neutrino flux \cite{Murase:2022dog}, and relaxing assumptions ({\it e.g.}, with longer CR cooling timescales and/or softer CR spectra) will make the limits stronger. 
Future multimessenger observations and astrophysical modeling will allow us to better understand the sources and reduce uncertainties, and the resulting limits on the DM-proton and the DM-electron cross section will become more stringent and robust. Understanding acceleration mechanisms will also enable us to compare $\tau_{{\rm DM}-i}^{\rm el}$ to the acceleration timescale for placing constraints.   

Multimessenger observations of neutrino sources have been proposed to study DM interactions with photons \cite{Ferrer:2022kei} and neutrinos \cite{Arg_elles_2017,Kelly:2018tyg,Choi:2019ixb,Murase_2019,Alvey_2019,Eskenasy:2022aup,Cline:2022qld,Ferrer:2022kei,Cline:2023tkp}, as well as historically-investigated annihilating or decaying signatures.
Now there is accumulating evidence that AGN can accelerate CRs to TeV--PeV energies. We demonstrate that high-energy particle emission from AGN provides us with a powerful probe of DM scatterings with protons and electrons through CR interactions.

\vspace{1mm}

{\it Acknowledgements---.}We are very grateful to Francesca Capel, Mar Císcar, Francesc Ferrer, Alejandro Ibarra, Walter Winter, Chengchao Yuan, and Bing Zhang for useful discussions. 
KM thanks the Topical Workshop: NGC 1068 as cosmic laboratory sponsored by SFB1258 and Cluster of Excellence ORIGINS. 
GH is supported by the Collaborative Research Center SFB1258 and by the Deutsche Forschungsgemeinschaft (DFG, German Research Foundation) under Germany's Excellence Strategy - EXC-2094 - 390783311. 
KM is supported by the NSF Grant No.~AST-1908689, No.~AST-2108466 and No.~AST-2108467, and KAKENHI No.~20H01901 and No.~20H05852.

\bibliography{References}    

\clearpage

\appendix

\section{DM spike formation and uncertainties}
Our method, which enables us to probe DM-proton and DM-electron scatterings through CR cooling in AGN, are insensitive to uncertainties in CR spectra as long as we rely on energetics arguments of AGN. The major uncertainty comes from the DM density profile, and it is worthwhile to justify the DM spike formation and discuss related uncertainties.

The DM spike formation is expected when a SMBH forms in the center of the DM distribution. For AGN, the offset of the SMBH with respect to the center of the host galaxy is typically less than $\sim1-100$~pc~\cite{Bartlett:2020uqa}, in which the DM spike may grow while the central SMBH coevolves with the galaxy. The major merger process may disrupt the DM spike, but NGC 1068 is morphologically similar to a normal spiral galaxy even though a minor merger happened in the past, and NGC 1068 is effectively regarded as a nonmerger object~\cite{Yamada+21}. TXS 0506+056 is a blazar, and the merger history of the host galaxy is more uncertain. Given that these AGN host SMBHs in the center and coevolve without recent major merger events, the DM spike formation is naturally expected if the adiabatic growth is realized. 
The Salpeter timescale characterizes the growth rate of black holes, and is given by $t_S=M_{\mathrm{BH}} / \dot{M}_{\mathrm{Edd}}\simeq4.5\times10^7$~yr, where $\dot{M}_{\mathrm{Edd}}$ is the Eddington accretion rate. This can be calculated as $\dot{M}_{\rm Edd} = L_{\rm Edd}/(\eta_{\rm rad} c^2)$, where $\eta_{\rm rad}\approx0.1$ is the radiative efficiency, and we note that the realistic mass accretion rate should be lower than $\dot{M}_{\mathrm{Edd}}$ for sub-Eddington accretion objects. The Salpeter timescale is to be compared with the dynamical timescale in the region of the black hole dominance, $t_{\rm dyn}=G M_{\mathrm{BH}} / \sigma^3$, where $\sigma$ is the velocity dispersion of the stars and DM particles \cite{Sigurdsson:2003wu}.

For NGC 1068, the SMBH growth timescale can be estimated from the bolometric luminosity. The black hole growth timescale is $t_{\rm grow}^{\rm NGC}=t_{S}\lambda_{\rm Edd}^{-1}\sim10^8$~yr, since the accretion rate is $\dot{M}_{\mathrm{acc}} = (L_{\rm bol}/\eta_{\rm rad} c^2)$, and the Eddington parameter $\lambda_{\rm Edd} \equiv L_{\mathrm{bol}} / L_{\mathrm{Edd}} \sim 0.5$~\cite{Murase:2022dog}. The dynamical timescale can be estimated from the black hole mass and velocity dispersion of stars, which is $\sigma_{\rm NGC} \sim 100$ km/s \cite{richstone}. We find $t_{\rm dyn}^{\rm NGC} \sim 10^6$~yr, so NGC 1068 is well within the adiabatic growth regime. Moreover, NGC 1068 is a jet-quiet AGN which would not cause additional disrupting processes that may alter the structure of the DM spike. Thus, we conclude that it is reasonable for NGC 1068 has a DM spike.

The black hole mass of TXS 0506+056 is much more uncertain than for NGC 1068, and it is difficult to estimate  $t_{\rm dyn}^{\rm TXS}$. On the other hand, TXS 0506+056 is usually classified as a BL Lac object, in which the accretion rate is expected to be much lower than $\dot{M}_{\mathrm{Edd}}$. Thus, we still find that the adiabatic growth holds within these uncertainties. We note that TXS 0506+056 is a blazar with a relativistic jet, which may affect the structure of the spike especially if a DM-proton coupling is present. This uncertainty is covered in this Appendix, and in our model (II), where we considered a spike substantially depleted by self-annihilations of DM particles.

\begin{figure*}[t!]
		\centering
		\includegraphics[width=0.44\textwidth]{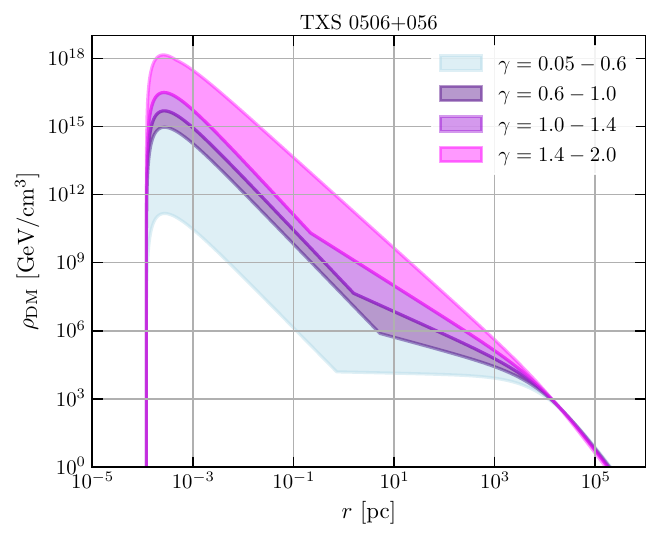}
		\includegraphics[width=0.46\textwidth]{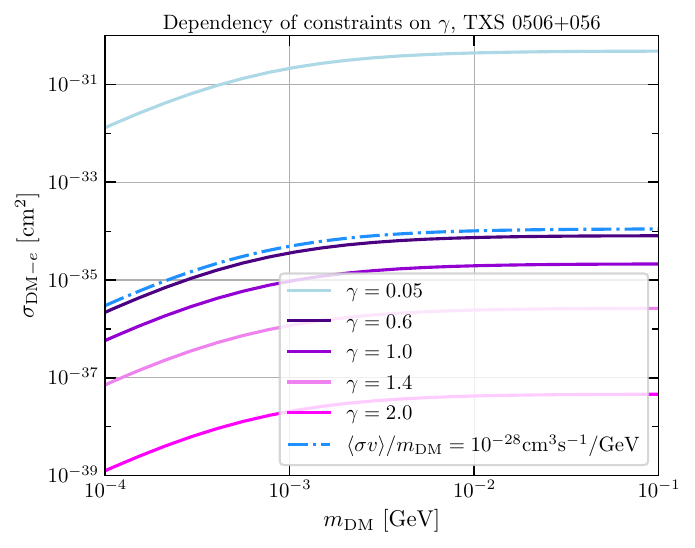}
        \includegraphics[width=0.46\textwidth]{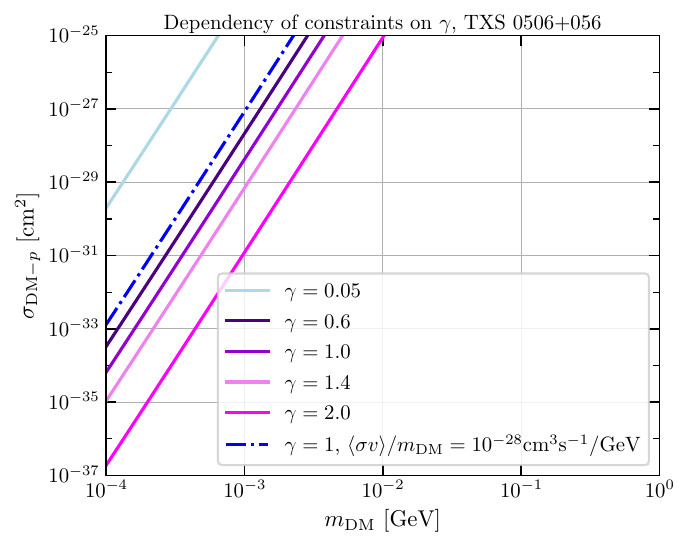}
		\caption{{\it Upper panels}: DM distribution in TXS 0506+056, for different values of $\gamma$. {\it Lower panels}: Upper limits on the sub-GeV DM-electron and DM-proton cross section from TXS 0506+056, for different values of $\gamma$.}
		\label{fig:DependencyGamma_TXS}
\end{figure*}

\begin{figure*}[t!]
		\centering
        \includegraphics[width=0.48\textwidth]{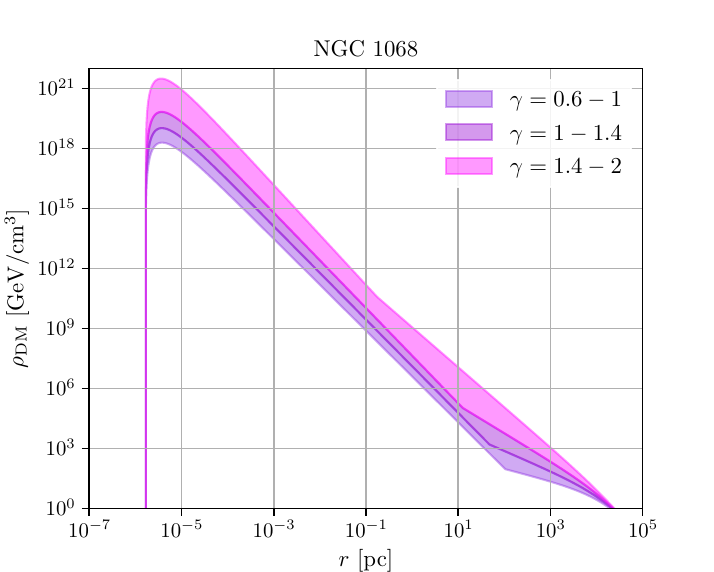}
        \includegraphics[width=0.46\textwidth]{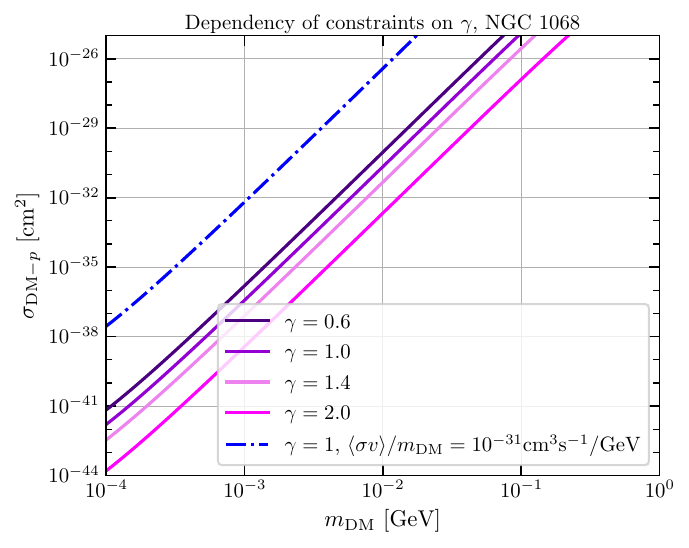}
		\caption{{\it Left panel}: DM distribution in NGC 1068, for different values of $\gamma$. {\it Right panel}: Upper limits on the DM-proton cross section from NGC 1068, for different values of $\gamma$.}
		\label{fig:DependencyGamma_NGC}
\end{figure*}

Assuming that the adiabatic growth holds, we discuss the impacts of uncertainties. In the main text, we derived constraints on the DM-proton and DM-electron cross section for different values of the pre-existing DM profile slope index. 
We consider NFW-like profiles with slope indices in the range $\gamma=0.05-2$ for TXS 0506+056 and $\gamma=0.6-2$ for NGC 1068, and normalize all profiles such that the total mass of the DM halo of TXS 0506+056 is bound by $M_{\rm DM} \lesssim 10^{13} M_{\odot}$, and the total mass of the DM halo of NGC 1068 is bound by $M_{\rm DM} \lesssim 10^{12} M_{\odot}$, consistent with expected relations between SMBH masses and DM halo masses in galaxies \cite{DiMatteo:2003zx}.
In Fig.~\ref{fig:DependencyGamma_TXS} and Fig.~\ref{fig:DependencyGamma_NGC}, we show the DM profiles in TXS 0506+056 and NGC 1068 for different values of $\gamma$. We also show in the Figures upper limits on the DM-electron cross and DM-proton cross section found for those profiles. The constraints can vary 5 to 6 orders of magnitude depending on the value of $\gamma$, although only vary within 1 order of magnitude in the range $\gamma=0.6-1.4$, which is favored by some simulations \cite{Jing:1999ir, Reed:2003hp}. 
Shallower profiles might arise \textit{e.g.,} due to the gravitational scattering of DM with stars \cite{Merritt:2003qk, Gnedin:2003rj}, with $\gamma_{\rm sp}=1.5$, among other possibilities \cite{Merritt:2006mt, Ullio:2001fb}. Dynamical constraints on the existence of spikes have been derived for some sources, disfavoring spikes with $\gamma>1$ \cite{Lacroix:2018zmg, Shen:2023kkm, Chan:2022gqd, Alachkar:2022qdt}. Further, evidence for the presence of parsec scale dark matter spikes around black holes has been suggested in \cite{Chan:2022gqd, chan2024robust}.


\clearpage

\section{Implications for a concrete model of DM-proton interactions}

In the main text, we derived generic constraints on DM-proton scatterings mediated by a heavy scalar mediator $m_{\phi} \gg q$. Here, we derive constraints in a concrete model with a hadrophilic scalar mediator, studying the validity of our constraints at different mediator masses. This allows for a more proper comparison of our constraints with complementary ones, and to connect our phenomenological probe with the parameter space favored by freeze-in DM production in the early universe \cite{Elor:2021swj, Bhattiprolu:2022sdd}. Besides freeze-in, we note that this region of parameter space is also allowed for some non-trivial thermal dark matter production mechanisms \cite{Smirnov:2020zwf,Parikh:2023qtk}.

\begin{figure}[H]
		\centering
		\includegraphics[width=0.46\textwidth]{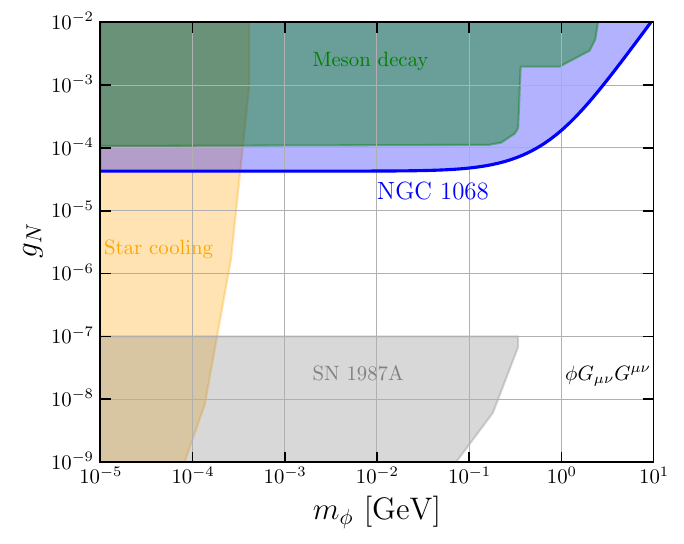}
        \caption{Constraints in a hadrophilic scalar mediator from NGC 1068 (Model I) for $m_{\chi}= 1$ MeV and $g_{\chi}=\pi$. The shaded blue region is ruled out by multimessenger observations from NGC 1068. We also show complementary constraints from meson decays, star cooling and SN 1987A \cite{Knapen:2017xzo}. Complementary constraints correspond to the scenario where the scalar mediator couples to nucleons via gluons with operator $\phi G_{\mu \nu}G^{\mu \nu}$.
}
		\label{fig:bounds_gn}
\end{figure}

If the interaction between the DM and CRs is mediated by a scalar with mass comparable or smaller to the momentum transfer of the process, the differential cross section from Eq. \ref{eq:differentialcrosssection} is corrected by
\begin{equation}
\frac{d \sigma_{\mathrm{DM}-\mathrm{CR} i}^{\phi}}{d T_{\mathrm{DM}}}=\frac{m_{\phi}^4}{\left(q^2+m_{\phi}^2\right)^2}\frac{d \sigma_{\mathrm{DM}-\mathrm{CR} i}}{d T_{\mathrm{DM}}} ,
\end{equation}
where $m_{\phi}$ is the mass of the scalar particle. The differential cross section is linearly proportional to the non-relativistic DM-nucleon scattering cross section, which can be related to the couplings of the underlying theory. We assume that the scalar $\phi$ couples to a DM fermion $\chi$ and nucleons $N$ via
\begin{equation}
\mathcal{L} \supset-m_\chi \bar{\chi} \chi-g_N \phi \bar{N} N-g_{\chi} \phi \bar{\chi} \chi .
\end{equation}
The non-relativistic DM-nucleon scattering cross section then reads
\begin{equation}
\sigma_{\mathrm{DM}-N}=\frac{g_\chi^2 g_N^2 \mu_{\chi-N}^2}{\pi m_{\phi}^4} .
\end{equation}
With these ingredients, we can derive constraints on the parameter space of $g_{N}-m_{\phi}$ and compare with complementary constraints. Our results are shown in Fig.~\ref{fig:bounds_gn}.

As can be appreciated in the Fig.~\ref{fig:bounds_gn}, CR cooling may allow to probe a new region of the parameter space of hadrophilic mediators unconstrained by other astrophysical and laboratory probes. We note that this is the case when the scalar mediators couples to nucleons via gluon operators. In models where the scalar couples directly to quarks, stronger constraints may apply \cite{Knapen:2017xzo}.

\end{document}